\documentclass{moriond}
\usepackage{amsmath}

\usepackage[numbers]{natbib}

\def\be{\begin{equation}}
\def\ee{\end{equation}}
\def\bea{\begin{eqnarray}}
\def\eea{\end{eqnarray}}

\newcommand{\BlackHat}{{\sc BlackHat}}
\newcommand{\SHERPA}{SHERPA}
\newcommand{\Comix}{COMIX}
\newcommand{\ROOT}{{\sc Root}}

\def\Wj{$W\,\!+\,1$}
\def\Wjj{$W\,\!+\,2$}
\def\Wjjj{$W\,\!+\,3$}
\def\Wjjjj{$W\,\!+\,4$}
\def\Wjjjjj{$W\,\!+\,5$}
\def\Wjjjjjj{$W\,\!+\,6$}
\def\Wmjjj{$W^-\,\!+\,3$}

\def\Wmjjjjj{$W^-\,\!+\,5$}
\def\Wmjjjjjj{$W^-\,\!+\,6$}
\def\Wpjjjjjj{$W^+\,\!+\,6$}
\def\WZF{$W,Z\,\!+\,\le4$}
\def\Wjn{$W\,\!+\,n$}
\def\Wjnm{$W\,\!+\,(n\!-\!1)$}
\def\Wmjn{$W^-\,\!+\,n$}
\def\Wmjnm{$W^-\,\!+\,(n\!-\!1)$}
\def\Wpjn{$W^+\,\!+\,n$}
\def\Wpjnm{$W^+\,\!+\,(n\!-\!1)$}

\def\murf{\mu_{\rm R,F}}

\def\kT{k_{\rm T}}
\def\pT{p_{\rm T}}
\def\HTjet{H_{\rm T}^{\rm jets}}
\def\HTmax{H_{\rm T}^{\rm max}}
\def\pTmin{p_{\rm T}^{\rm min}}

\newif\ifpreprint
\preprinttrue

\ifpreprint
\newcommand{\Photo}{}
\else
\newcommand{\Photo}{\includegraphics[height=35mm]{David-Kosower.jpg}}
\fi

\def\spacer{\hfil\hskip 10mm\hfil}
\begin{document}
\vspace*{4cm}
\title{
\ifpreprint
\hbox to\linewidth{\rm\small
UCLA-14-TEP-103\spacer
SLAC--PUB--15969\spacer
IPhT--T14/104\spacer
IPPP/14/069
\break}
\hbox{$\null$\break}
\fi
Universality in $W$+Multijet Production}

\author{Z.~Bern${}^a$, L.~J.~Dixon${}^b$, F.~Febres Cordero${}^c$, 
S.~H{\" o}che${}^b$, H.~Ita${}^{d}$, D.~A.~Kosower${}^e$, 
D.~Ma\^{\i}tre${}^{f}$ and K.~J.~Ozeren${}^a$
\\
$\null$
\\
${}^a$Department of Physics and Astronomy, UCLA, Los Angeles, CA
90095-1547, USA \\
${}^b$SLAC National Accelerator Laboratory, Stanford University,
             Stanford, CA 94309, USA \\
${}^c$Departamento de F\'{\i}sica, Universidad Sim\'on Bol\'{\i}var, 
 Caracas 1080A, Venezuela\\
${}^d${Physikalisches Institut, Albert-Ludwigs-Universit\"at Freiburg, D--79104 Freiburg, Germany}\\
${}^e$Institut de Physique Th\'eorique, CEA--Saclay,
          F--91191 Gif-sur-Yvette cedex, France\\
${}^f$Department of Physics, University of Durham, Durham DH1 3LE, UK\\
}

\address{\BlackHat{} \textrm{Collaboration}\vspace*{2mm}}

\maketitle\abstracts{
We study $W$-boson production accompanied
by multiple jets at 7~TeV at the LHC.
We study the jet-production ratio, of total cross sections for \Wjn- to \Wjnm-jet
production, and the ratio of distributions in the total
transverse hadronic jet energy $\HTjet$.
We use the ratios to extrapolate the total cross section,
and the differential distribution in $\HTjet$, to
\Wjjjjjj-jet production.
We use the \BlackHat{} software library
in conjunction with \SHERPA{} to perform the 
computations.
}

\section{Introduction}

The search for physics beyond the Standard Model relies on
quantitative theoretical calculations of known-physics backgrounds.
Uncovering signals of new physics requires a good quantitative
understanding of the backgrounds as well as the corresponding
theoretical uncertainties.  The challenge of performing the required
theoretical calculations increases with the increasing jet
multiplicities used in many search strategies.  We are thus encouraged
to look for features of the relevant Standard-Model processes that
can simplify calculations at higher multiplicities.

In this contribution, we study one of the benchmark Standard-Model processes,
production of a $W$ electroweak vector boson accompanied by multiple jets.
The short-distance matrix element can be computed systematically in
perturbative QCD.  A leading-order (LO) calculation, however, leaves a strong
dependence on the renormalization and factorization scales introduced in
order to define the coupling $\alpha_s$ and the parton distribution functions
(PDFs).  This unphysical dependence becomes stronger with an increasing
number of accompanying jets.  A next-to-leading order (NLO) calculation is
required to obtain a quantitatively reliable prediction.  We
expect such predictions to be accurate to 10--15\%.  Future improvements to
experimental uncertainties will demand that theorists go to yet higher order in
the perturbative expansion.

With the \BlackHat{} software library~\cite{BlackHatI}, building on the progress
in NLO calculations in recent years~\cite{UnitarityMethod, DDimUnitarity, Zqqgg, NewUnitarity,
  BCFW, GenHel, OPP, Forde, Badger}, we are able to perform high-multiplicity
calculations.  The software library supplies one-loop amplitudes,
using on-shell methods.   In this approach, the amplitude is written
as a sum over known integrals,
\begin{equation}
\textrm{Amplitude} = \sum_{j\in\rm{Basis}} c_j \textrm{Int}_j
+ \textrm{Rational}\,,
\end{equation}
where the integrals are the usual box, triangle, and bubble one-loop integrals.
The integrals coefficients $c_j$ as well as the additional rational
terms are rational functions of {\it spinor\/} variables.  They are
built out of Lorentz-invariant bilinear products of spinors.
The external momenta
are quadratic functions of these spinor variables, and Lorentz invariants
are in turn quadratic functions of Lorentz-invariant spinor products.
The \BlackHat{} library embodies a numerical implementation
of on-shell methods, and the unitarity method in particular.  The NLO
calculation as a whole, including phase-space
generation, multichannel integration, and 
subprocess management, is managed by \SHERPA{}~\cite{Sherpa}.  In addition, the
real-emission contributions, with subtractions according to the
Catani--Seymour scheme~\cite{CataniSeymour}
as modified by an $\alpha_{\rm dipole}$ cut-off
parameter~\cite{Nagy}, 
are produced using the \Comix{} library~\cite{Comix} (itself a part
of \SHERPA).

\section{$n$-Tuples}

While high-multiplicity calculations have become feasible with this
technology, they are still complex and do require considerable amounts
of computer time.  This is particularly true for 
the calculation of \Wjjjj- and \Wjjjjj-jet
production.  There are many different contributions --- different
subprocesses, leading color vs.{} subleading color, etc.{} --- to
track and manage, so that
running a calculation requires much more human effort as well as computer
time.   With a traditional set up, the whole calculation would need
to be re-run to study scale dependence or to compute a new distribution.
Even worse, the whole calculation would have to be rerun many times
in order to estimate the uncertainty due to PDFs.

The computational effort is overwhelmingly dominated by the 
squared matrix elements, whether virtual or real-emission.  The
computation of additional observables or differential distributions
is relatively cheap.  The same would be true of results for different
choices of renormalization or factorization scales $\murf$,
and for different PDFs, were we to record the terms within the 
matrix element that have different dependences on these quantities.

These considerations motivate us to recycle the matrix elements rather
than recomputing them.  We compute the matrix elements once, saving
phase-space configurations along with the weights, split up into the
coefficients of the different functional dependences on $\murf$ and
the PDFs~\cite{NTuples}.  We save this information as \ROOT{} $n$-tuple files.  Different types
of contributions to an NLO cross section --- the Born terms, virtual
corrections, subtracted real-emission corrections, and integrated-subtraction
terms --- are saved in separate files.  Each analysis --- computing
a new distribution, or assessing the scale dependence --- can then
be done with a lightweight code (in C++, \ROOT{}, {\sf Python\/},$\ldots$).

As a bonus, these $n$-tuples can be distributed to experimenters, 
who can perform their own analyses.  The only real restriction is
to a preselected set of jet algorithms and to a minimal value
of the jet $\pTmin$ cut.  The $n$-tuples we have
produced to date allow for the use of common LHC algorithms 
(SISCone, $\kT$, and anti-$\kT$, with $R=0.4,0.5,0.6,0.7$).
The remaining cuts --- jet, lepton, or photon minimum $\pT$ cuts, rapidity
cuts --- can be tightened, or additional cuts
 can be introduced.  Of course, if the additional
cuts are very tight, the $n$-tuple statistics of a calculation will suffer, and
it would be better to create a special set of $n$-tuples for the tighter cuts.  We have
made $n$-tuples for \WZF~jets and for four-jet production available,
and diphoton+dijet $n$-tuples (both for generic cuts and a special
set for VBF cuts) will follow soon.

\section{Jet-Production Ratios}

The transverse momentum distributions in \Wjjjj- and \Wjjjjj-jet 
production are shown in fig.~2 in ref.~\cite{W4j} and fig.~6 in ref.~\cite{W5j} respectively.
The calculations show greatly reduced scale dependence at NLO
compared to LO.  The NLO corrections soften the distribution for
all jets except the softest jet.  Successive jet distributions
become steeper and steeper.  This can be understood as reflecting
the greater increase of the partonic center-of-mass invariant $\hat s$ when increasing (say) the fifth
jet's $\pT$ as opposed to the third jet's, with all jets ordered
in decreasing $\pT$.  The increase in $\hat s$ in turn decreases
the matrix element as well as the PDFs.  Other than the increasingly
steep fall, distributions for the third-softest jet and softer
jets resemble each other.  This raises the question of whether
one can think of these jets as `generic', and whether one can
find patterns in \Wjn-jet production.

In order to examine patterns in $W$+multijet production in greater detail,
it is helpful to examine ratios of observable quantities in
\Wjn-jet production to that in \Wjnm-jet production.  Such ratios
should be less sensitive to experimental uncertainties: the luminosity
uncertainty should cancel, and the jet-energy scale dependence will be
lessened.  They should also be less sensitive to theoretical uncertainties:
though it is hard to quantify, the scale sensitivity should diminish;
and the dependence on the PDFs should decrease as well.

At low multiplicities, however, we expect deviations from generic behavior.
In \Wj-jet production, some subprocesses are missing at LO; and even
at NLO, there are strong kinematic restrictions on the phase space
of jets.  In \Wjj-jet production, there are strong kinematic restrictions
on phase space (the $W$ boson cannot be near the leading jet, for example),
which only start being relaxed at NLO.  In the $W$ $\pT$ distribution,
this is reflected in large NLO corrections beyond low $\pT$
in the \Wjjj/\Wjj-jet ratio,
while corrections to the \Wjn/\Wjnm-jet ratio for $n\ge 4$ are more modest.

\begin{figure}
\centerline{\includegraphics[width=0.7\linewidth]{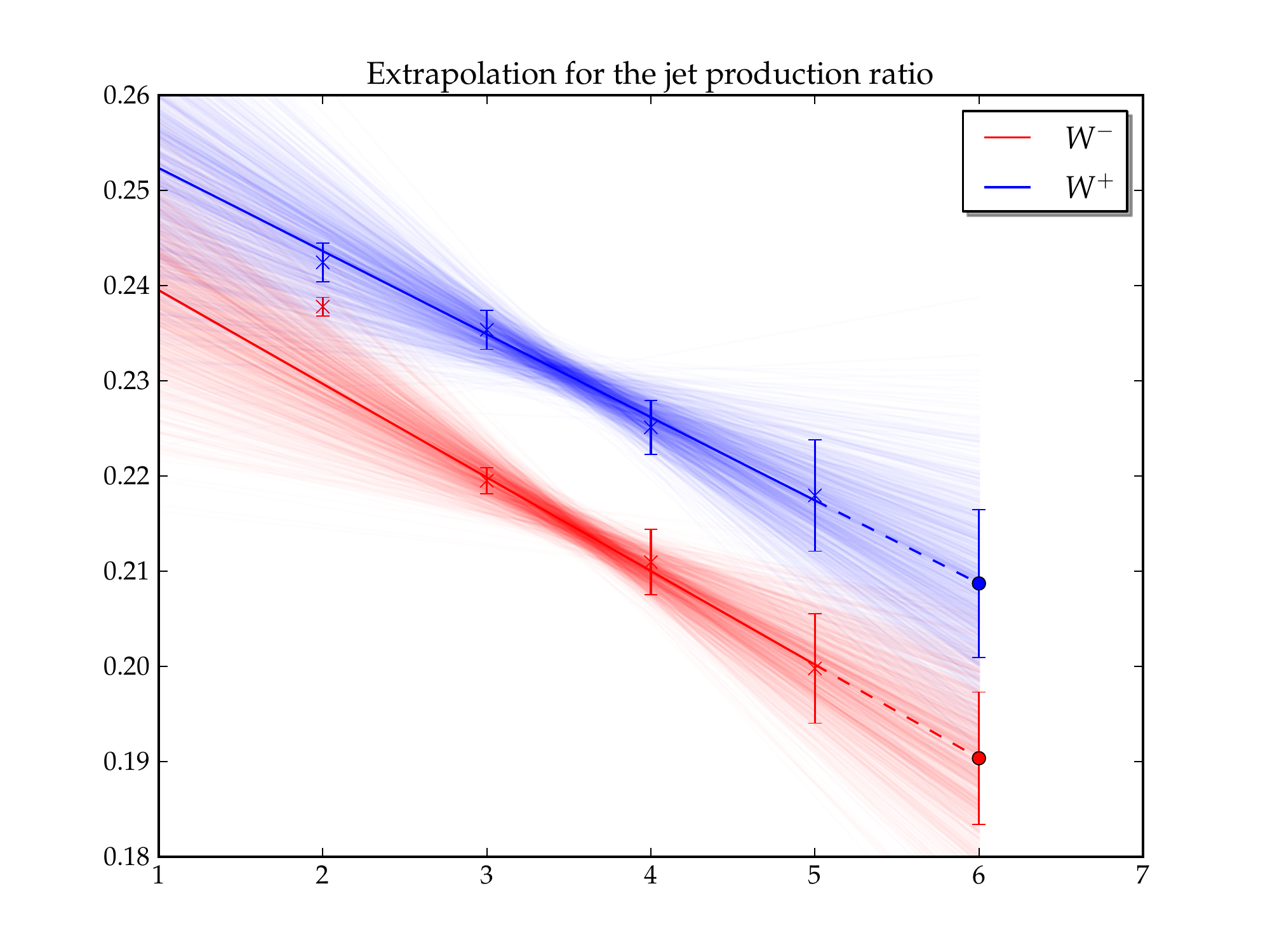}}
\caption[]{Extrapolations of the ratio of total cross sections at NLO.  The lower (red)
line shows the extrapolation for the \Wmjn/\Wmjnm ratio, and the upper (blue) line
for the \Wpjn/\Wpjnm ratio.  The fainter lines illustrate the uncertainty
envelope based on the statistical uncertainty of each underlying cross
section.}
\label{fig:TotalCrossSectionExtrapolation}
\end{figure}

The simplest pattern we can seek is in the 
ratio of total cross sections subject to standard jet cuts. 
(The cuts we use are given in ref.~\cite{W5j}.) In 
fig.~\ref{fig:TotalCrossSectionExtrapolation}, we show a linear fit
to the cross-section ratios.  The fit to the \Wjjj/\Wjj-jet, \Wjjjj/\Wjjj-jet,
and \Wjjjjj/\Wjjjj-jet ratios is very good.  The \Wjjjjj-jet calculation is
needed to make this assessment meaningful; were it absent, we might even
be misled into including the \Wjj/\Wj-jet ratio.  As can be seen in the
figure, and as expected, that ratio (at least for $W^-$) is quite different
from what would be expected from the fit.  The linear fit allows us to
predict the \Wjjjjjj/\Wjjjjj-jet ratio, and that in turn, allows us
give a prediction~\cite{W5j} for the \Wjjjjjj-jet production cross section,
\begin{equation}
\begin{aligned}
\textrm{\Wmjjjjjj{} jets:} &\quad\! 0.15\pm 0.01~{\rm pb}\,,\\
\textrm{\Wpjjjjjj{} jets:} &\quad\! 0.30\pm 0.03~{\rm pb}\,,\\
\end{aligned}
\label{eqn:W+SixJets}
\end{equation}
where the uncertainty estimates include only statistical uncertainties.

\section{Extrapolating the $\HTjet$ Distribution}

\begin{figure}
\centerline{\includegraphics[width=0.5\linewidth]{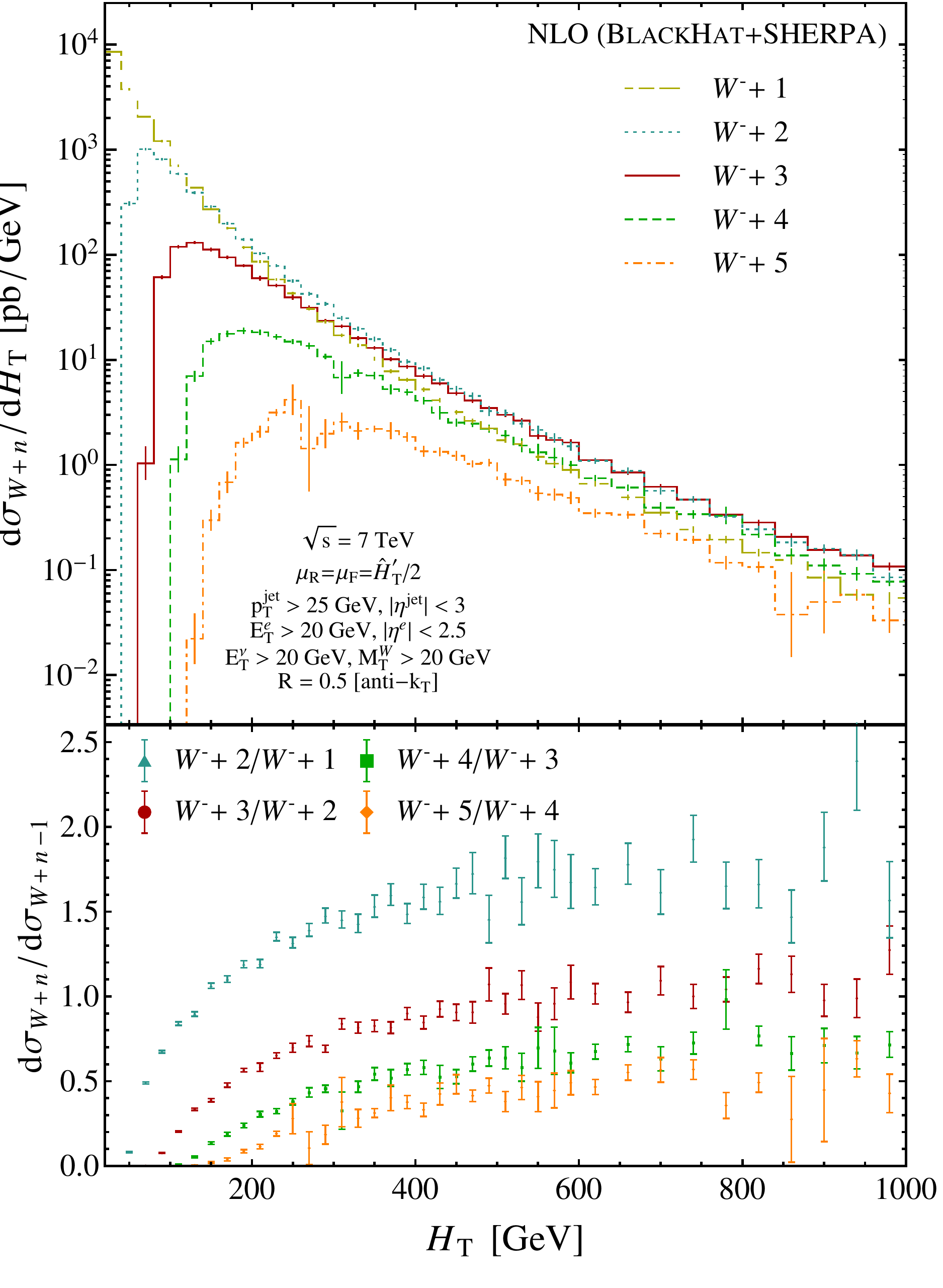}
\hfil
\raisebox{2.1mm}{\includegraphics[width=0.5\linewidth]{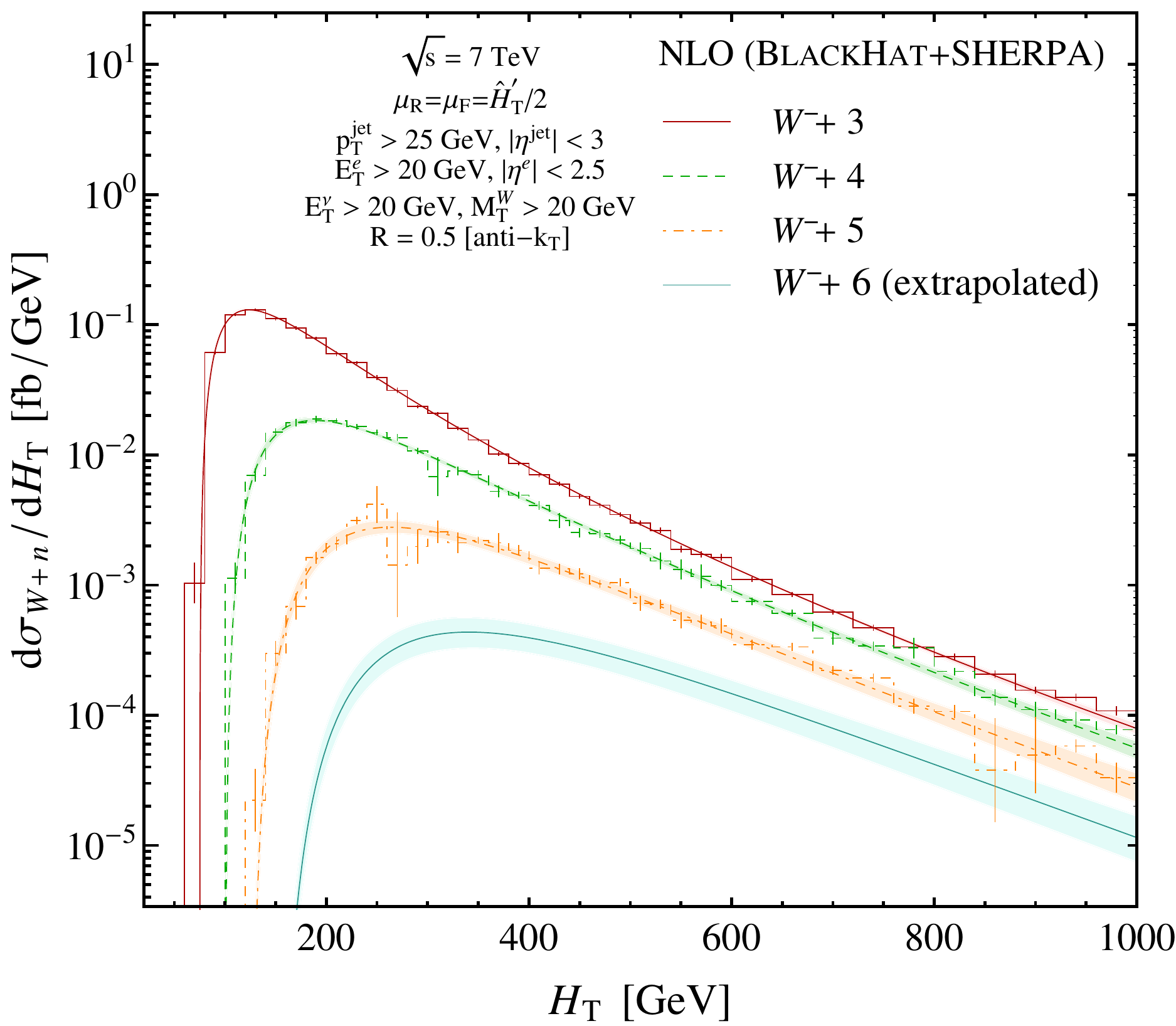}}}
\caption[]{(a) In the upper panel at left, the computed $\HTjet$ distributions at NLO for $W^-$ production 
accompanied by up to five jets; in the lower panel, ratios of these distributions (b) At
right, the computed $\HTjet$ distributions
at NLO
for $W^-$ production with three to five accompanying jets, compared
with an ratio-based parametrization, along with the expolation to 
\Wmjjjjjj-jet production.}
\label{fig:HTjets}
\end{figure}

The total transverse energy in jets, $\HTjet$, is a good probe into
potential new physics at the very highest center-of-mass energies accessible
to the LHC.  We have computed this distribution for a $W$ boson accompanied
by up to five jets.  The results are shown in the upper panel of
fig.~\ref{fig:HTjets}(a).  Let us try to predict the distribution for
\Wjjjjjj-jet production by extrapolating those results.  
The $\HTjet$ distributions have
a threshold, due to the minimum jet transverse momentum.  Combined
with the decrease towards larger $\HTjet$, due to the decreasing matrix 
elements and the falling parton distributions, this leads to the
appearance of a peak in the distribution.  The threshold, and hence the
peak locations, are different for different numbers of jets.  This
makes a simple extrapolation at each different value of $\HTjet$ problematic.
Instead, we seek to fit a functional form.  

At small $\HTjet$, we might expect the integral we are evaluating for $n$ jets to
have the following form,
\begin{equation}
\biggl(\int\frac{dE}{E} g(E)\biggr)^m\,,
\end{equation}
where $g(E)$ is slowly varying, and where $m<n$ because
not all jets can be soft.  This form suggests a functional form of $\ln^\tau \rho$ to include as a factor,
where $\rho = \HTjet/(n \pTmin)$.  We take $\tau$ as a fit parameter.
At very large $\HTjet$ (well to the right of the plots in 
fig.~\ref{fig:HTjets}), phase space becomes constrained, suggesting
a factor like $(1-\HTjet/\HTmax)^\gamma$, where $\HTmax \simeq 7$~TeV,
and $\gamma$ is a fit parameter.  We have previously seen~\cite{TevZ} that such a factor is
appropriate for the Tevatron, where it is more noticeable because of the
lower center-of-mass energy.

This suggests the following fit form,
\begin{equation}
\frac{d\sigma^{W+n}}{d\HTjet} = a_s^n N_n \ln^{\tau_n} \rho_n 
(1-\HTjet/\HTmax)^{\gamma_n}\,,
\end{equation}
where $N_n$ is a normalization, and 
$a_s\equiv \alpha_s(\HTjet) N_c/\pi$.  Using this form gives a poor fit to the
$\HTjet$ distributions themselves, but an excellent fit for
the ratios of these distributions shown in the bottom panel of fig.~\ref{fig:HTjets}(a).  We can then fit for the values of
the $\tau$ and $\gamma$ parameters; once again, we get a good linear fit.
Using either a fit with additional
parameters to the \Wjj-jet $\HTjet$ distribution, or the numerical
values of this distribution directly, we can then use predicted values of
the $\tau$ and $\gamma$ parameters to obtain predictions for the
$\HTjet$ distribution in \Wjn-jet production.  
The normalization $N_n$ is related
to the total cross section; we can solve for it by integrating the fit
form using the extrapolated values of $\tau$ and $\gamma$, and comparing
with the extrapolated value of the total cross section as 
in eq.~(\ref{eqn:W+SixJets}).  We can cross check this procedure by
comparing the `predicted' curves for \Wmjjj- through \Wmjjjjj-jet production
with the directly computed values; this comparison is shown in
fig.~\ref{fig:HTjets}(b).  The predicted $\HTjet$ distribution in
\Wmjjjjjj-jet production is shown in the same figure.

\section{Conclusions}

The production of an electroweak vector boson accompanied by multiple
jets is an important Standard-Model process at the LHC.  We have studied
this process in a wide range of kinematic regimes, and with a varying
number of accompanying jets.  We see indications of simple and regular
behavior in NLO calculations of
 cross sections and distributions beginning with three
accompanying jets.  We have used this behavior to predict the total
cross section and a key distribution in \Wjjjjjj-jet production.

\section*{Acknowledgments}

This research was
supported by the US Department of Energy under contracts
DE--AC02--76SF00515 and DE--SC0009937.  DAK’s research is
supported by the European Research Council under Advanced Investigator
Grant ERC--AdG--228301.  DM's work was supported by the Research
Executive Agency (REA) of the European Union under the Grant Agreement
number PITN--GA--2010--264564 (LHCPhenoNet). SH's work was partly
supported by a grant from the US LHC Theory Initiative through NSF
contract PHY--0705682.  This research used resources of Academic
Technology Services at UCLA.


\end{document}